\documentclass[conference]{IEEEtran}
\IEEEoverridecommandlockouts
\usepackage{cite}
\usepackage{amsmath,amssymb,amsfonts}
\usepackage{graphicx}
\usepackage{textcomp}
\usepackage[dvipsnames]{xcolor} 
\usepackage{threeparttable}
\usepackage{array}
\usepackage{multirow}
\usepackage{booktabs}
\usepackage{algorithm}
\usepackage{algorithmic}
\usepackage{balance}
\usepackage{tabularx}
\usepackage{microtype}
\usepackage{enumitem}
\usepackage[caption=false,font=footnotesize]{subfig} 
\usepackage[a4paper, total={184mm,239mm}]{geometry}
\usepackage{tikz}
\DeclareRobustCommand\encircle[1]{\tikz[baseline=(char.base)]{\node[shape=circle,fill,inner sep=0.75pt] (char) {\textcolor{white}{#1}}}}

\usepackage{titlesec}
\titlespacing*{\section}{0pt}{1ex plus 0.6ex minus .2ex}{.4ex plus .1ex}
\titlespacing*{\subsection}{0pt}{1ex plus .1ex minus .0ex}{.1ex plus .0ex}
\titlespacing*{\subsubsection}{0pt}{1ex plus 1ex minus 1ex}{1ex plus 1ex}

\renewcommand{\thesubsubsection}{\arabic{subsubsection}}
\titleformat{\subsubsection}[runin]{\itshape}{\thesubsubsection)}{1em}{}
\titlespacing*{\subsubsection}{\parindent}{0pt}{*1}

\def\BibTeX{{\rm B\kern-.05em{\sc i\kern-.025em b}\kern-.08em
    T\kern-.1667em\lower.7ex\hbox{E}\kern-.125emX}}

\begin{document}

\title{CHIME: Chiplet-based Heterogeneous Near-Memory Acceleration for Edge Multimodal LLM Inference \vspace{-0.2cm}}

\author{
    \IEEEauthorblockN{
        Yanru Chen\textsuperscript{*1}, 
        Runyang Tian\textsuperscript{*1}, 
        Yue Pan\textsuperscript{1}, 
        Zheyu Li\textsuperscript{1}, 
        Weihong Xu\textsuperscript{2}, 
        Tajana Šimunić Rosing\textsuperscript{1}
    }

    \IEEEauthorblockA{
        \textsuperscript{1}University of California San Diego, La Jolla, CA, USA\\
    }

    \IEEEauthorblockA{
        \textsuperscript{2}Ecole Polytechnique Fédérale de Lausanne, Lausanne, Switzerland\\
    }

    \IEEEauthorblockA{
        \{yac054, r3tian, k4fan, yup014, zhl178, tajana\}@ucsd.edu; weihong.xu@epfl.ch
    }

}

\IEEEaftertitletext{\vspace{-1cm}}

\maketitle

\begin{abstract}
The proliferation of large language models (LLMs) is accelerating the integration of multimodal assistants into edge devices, where inference is executed under stringent latency and energy constraints, often exacerbated by intermittent connectivity. These challenges become particularly acute in the context of multimodal LLMs (MLLMs), as high-dimensional visual inputs are transformed into extensive token sequences, thereby inflating the key–value (KV) cache and imposing substantial data movement overheads to the LLM backbone.
To address these issues, we present CHIME, a chiplet-based heterogeneous near-memory acceleration for edge MLLMs inference. CHIME leverages the complementary strengths of integrated monolithic 3D (M3D) DRAM and RRAM chiplets: DRAM supplies low-latency bandwidth for attention, while RRAM offers dense, non-volatile storage for weights. This heterogeneous hardware is orchestrated by a co-designed mapping framework that executes fused kernels near data, minimizing cross-chiplet traffic to maximize effective bandwidth. On FastVLM (0.6B/1.7B) and MobileVLM (1.7B/3B), CHIME achieves up to $54\times$ speedup and up to $246\times$ better energy efficiency per inference as compared to the edge GPU NVIDIA Jetson Orin NX. It sustains $116.5-266.5$ token/J  compared to Jetson’s $0.7-1.1$ token/J. Furthermore, it delivers up to $69.2\times$ higher throughput than the state-of-the-art PIM accelerator FACIL. Compared to the M3D DRAM-only design, CHIME’s heterogeneous memory further improves energy efficiency by $7\%$ and performance by $2.4\times$.
\end{abstract}

\begin{IEEEkeywords}
on-device MLLMs, processing near-memory, heterogeneous memory chiplets, monolithic 3D, mapping framework
\end{IEEEkeywords}

\section{Introduction}

On-device AI is expanding with billions of edge devices, driven by demands for latency, offline availability, privacy, and strict energy constraints. Large language models (LLMs) are reshaping human-computer interaction, and multimodal LLMs (MLLMs) extend language models to vision and audio, including medical visual question answering (VQA) and report generation~\cite{qu2025mobileedgeintelligencelarge}. However, deployment on mobile platforms is often limited to small-batch inference, which exposes a critical performance bottleneck. MLLMs exacerbate the memory bottleneck, as multimodal inputs not only expand the Key-Value (KV) cache but also introduce significant cross-modal data transfers, increasing per-step memory traffic on already limited on-device bandwidth. Consequently, throughput is dictated by memory bandwidth, and energy consumption is dominated by off-chip data movement~\cite{li2025surveylargelanguagemodel}. 

% Existing approaches are insufficient: software optimizations like PagedAttention~\cite{10.1145/3600006.3613165} improve Key Value-cache (KV cache) allocation but fail to reduce the total bytes moved per token and therefore cannot address the bandwidth-dominated bottleneck.  Mobile SoCs are architecturally optimized for the convolutional reuse of vision workloads and thus are mismatched for the bandwidth-intensive access patterns of transformers. 

Existing approaches attempt to mitigate this bottleneck with software and system-level optimizations. While orchestrating heterogeneous computation across GPUs and neural accelerators improves throughput, it introduces a significant synchronization overhead, graph compilation cost and unpredictable power consumption~\cite{chen2025heterollm,xu2025_fast_npu}. Consequently, industry efforts have converged on mitigating data movement through techniques such as buffer compression, operator fusion, and low-precision quantization~\cite{wang2016_memsqueezer, apple2022_anetransformers,qualcomm2024_ondevice_npu, mediatek2024_genai_whitepaper}. These prior works, however, fail to overcome the physical von Neumann bottleneck between compute and memory. Thus, a paradigm shift in hardware architecture is required to unlock substantial on-device performance gains.

% These observations motivate a memory-centric substrate that raises effective bandwidth where data resides.

Processing-in-memory (PIM) has emerged as a promising approach to mitigate the memory wall and improve energy efficiency in LLMs by embedding computation directly within memory. Early systems largely targeted single-modal LLM. Newton proposed a DRAM-based PIM architecture that integrates multiply–accumulate (MAC) units directly into DRAM and reports an average $54\times$ speedup over a Titan V GPU on BERT transformer~\cite{he2020newton}. TransPIM augments HBM with lightweight computation units and token-based dataflow, delivering $115\times$ GPU speedup and up to $667\times$ energy efficiency~\cite{zhou2022transpim}. 

Recent systems extend toward multimodality. MulTCIM achieves $2.24\,\mu\mathrm{J}/\mathrm{Token}$ and $85.8$\,TOPS/W by exploiting hybrid sparsity with a modal-adaptive computing in memory network~\cite{tu2023multcim}; StreamDCIM adopts a tile-based reconfigurable macro with a mixed-stationary dataflow, improving throughput and energy by $1.28\times$ and $1.23\times$ over layer streaming~\cite{qin2025streamdcim}; PIM–GPU collaboration further shows that mapping framework-guided task partitioning can deliver up to $15\times$ end-to-end speedup over GPU-only baselines~\cite{ji2024real}. 

On-device MLLMs face a unique memory bottleneck: cross-modal attention demands high bandwidth, while multimodal inputs expand the KV cache, increasing storage pressure. While DRAM-centric systems struggle to meet both demands, we introduce CHIME, a chiplet-based near-memory design tailored for edge MLLMs. It integrates monolithic 3D (M3D) DRAM for latency-critical attention and connector kernels with M3D RRAM for high density weights and KV storage, under a mapping framework that places data and computation kernels to reduce movement, raise effective bandwidth, and respect RRAM endurance. Our core contributions are summarized below.

\begin{itemize}[leftmargin=*]
\item Efficient heterogeneous chiplet PIM architecture: A 2.5D advanced UCIe package links M3D DRAM for latency-critical kernels with M3D RRAM for energy-efficient storage.  

\item Hardware and software co-design: Our CHIME system manages weight and KV cache allocation, tiering, and migration, and schedules near-memory kernels to minimize cross-chiplet traffic and fully exploit heterogeneous memory.

\item Mapping framework for general MLLMs: The mapping framework maps operators to DRAM or RRAM by access patterns, schedules processing elements (PEs), special function processing elements (SFPEs) execution, and fuses kernels to keep activations local and cut transfers.

\item Speedup and energy efficiency: We compare CHIME with Jetson Orin NX on MobileVLM and FastVLM, reporting $31–54\times$ speedup and $113–246\times$ energy efficiency. Furthermore, CHIME achieves up to a $69.2\times$ higher throughput than the state-of-the-art (SOTA) PIM accelerator.
\end{itemize}

\section{Motivation and Background}

\subsection{MLLM Architecture}

\begin{figure}[ht]
    \centering
    \includegraphics[width=0.83\linewidth]{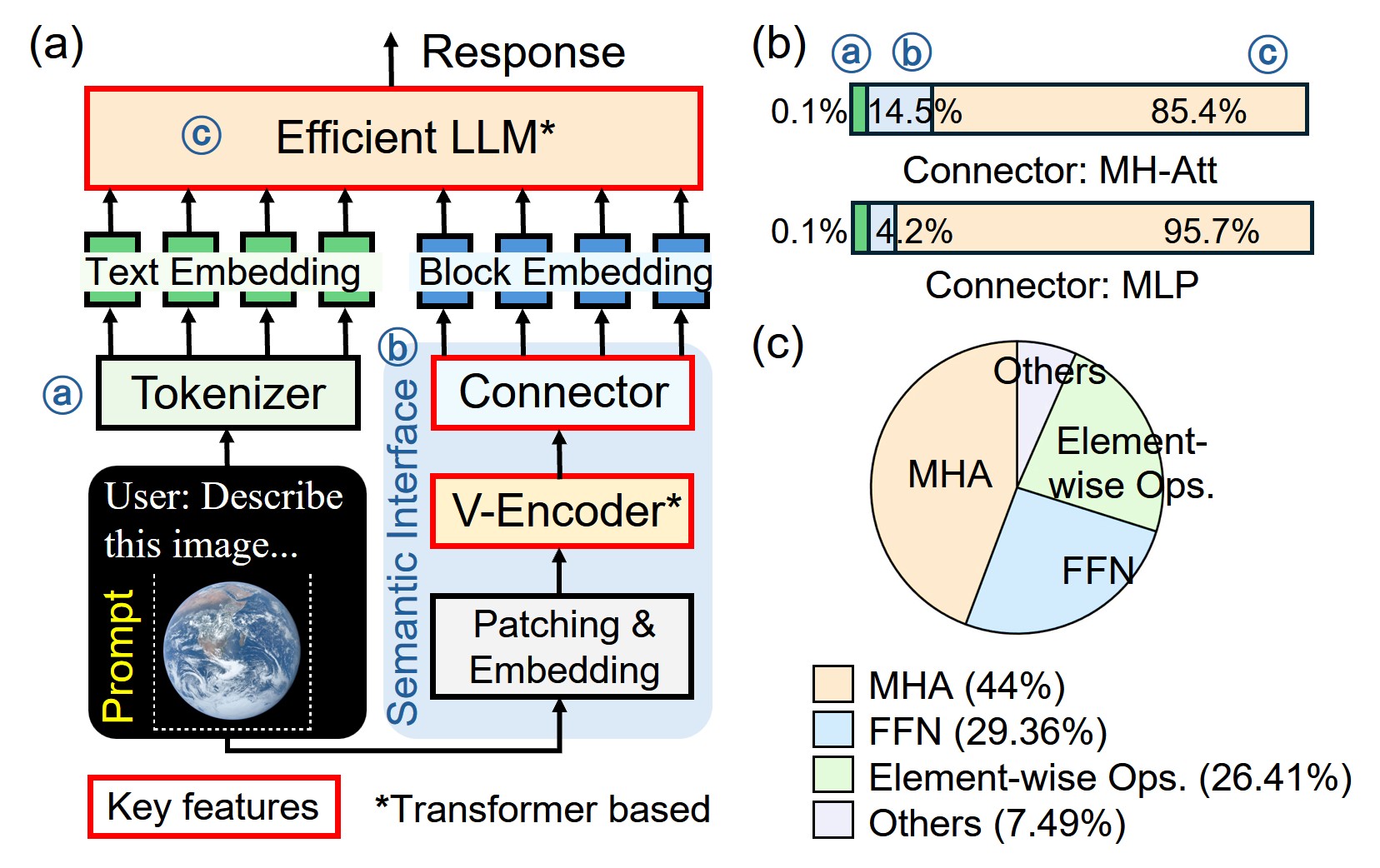}
    \caption{MLLM software architecture overview (a) MLLM with three key features (b) Execution time breakdown of MLLMs under different connectors (c) Execution time breakdown of the GPT-2 backbone on GPU~\cite{han2025sal}}
    \label{fig:fig1}
\end{figure}

MLLMs extend single-modal LLMs to reason over multi-modalities. As shown in Fig.~\ref{fig:fig1}(a), the architecture augments an LLM with a vision encoder and a connector. The encoder converts an image into feature embeddings, often with a vision transformer. The connector projects these embeddings into the language domain through multihead attention (MHA) projector or a multilayer perceptron (MLP) projector, producing pseudo-tokens. Together they form a semantic interface that transforms images into token sequences comparable to text, which are then fused and processed by the LLM backbone to generate responses. MLLMs break the linear scaling behavior of single-modal LLMs by inflating the KV cache through visual tokens and introducing irregular memory access and asynchronous compute through cross-modal fusion.

Profiling reveals that this semantic interface adds little computational load, while the LLM backbone remains the dominant cost. As Fig.~\ref{fig:fig1}(b) shows, the backbone accounts for 85.4\%–95.7\% of execution time, with the encoder and connector together consuming only 4.2\%–14.5\%. The dominant overhead is the transfer of large visual features from the memory hierarchy to the LLM backbone. Within the LLM backbone itself, execution is concentrated in transformer kernels, with MHA at 44\% and Feed-Forward Networks (FFN) at 29.36\% (Fig.~\ref{fig:fig1}(c)), followed by element-wise operations 26.41\%.

\subsection{Processing In Memory}

PIM is broadly classified into two types: processing-using-memory (PUM) and processing-near-memory (PNM)~\cite{pimprimer}. In DRAM, PUM is achieved by carefully timed multi-row activation with charge sharing, enabling batch row-wise bitwise and majority operations~\cite{ambit}. PNM augments DRAM stacks with digital compute near memory, transferring data through short internal links. High-bandwidth memory (HBM) offers a natural foundation for this model, featuring a stacked architecture with a base logic die for computation. 

% However, limited area and thermal headroom on the base logic die restrict near-memory computation and buffering in stacked DRAM systems~\cite{ndpsurvey}.

In resistive crossbars, PUM uses stateful logic for column-parallel Boolean primitives. RACER~\cite{racer} applies bit-pipelined execution across tiles to raise throughput. Analog crossbars support matrix–vector multiplication, but device variability, nonlinearity, and calibration overheads limit accuracy. Digital PUM avoids these non-idealities, preserving exact arithmetic for MLLM inference~\cite{pimprimer}. PNM designs pair RRAM arrays with digital logic, exploiting short-range access paths and lower write energy. We adopt PNM on M3D RRAM to balance precision, power, and area, preserving digital accuracy while M3D stacking raises bandwidth and capacity within thermal limits. Section~\ref{sec:M3D} details the integration.

\subsection{Monolithic 3D Devices}
\label{sec:M3D}

M3D sequentially builds multiple active layers on a single substrate and links them with nanometer-pitch, monolithic inter-tier vias (MIVs). For data-intensive applications, M3D integration is leveraged to develop PIM capable memories, using its fine-grain and low-latency vertical interconnects to move compute logic closer to or within memory and alleviate data movement costs~\cite{ramanathan2021cim3d}. Compared to through-silicon-via (TSV) stacking, MIVs provide far higher vertical interconnect density and lower capacitance and power, enabling tighter memory-compute coupling beyond 2D limits.

In this work, we focus on two complementary M3D devices. In DRAM, coarse TSV pitch limits vertical bandwidth in stacked memories such as HBM; replacing TSVs with dense MIVs in monolithic integration alleviates this bottleneck~\cite{felfel2020quantifying,dhananjay2021monolithic,pan2025stratum}. The vertical staircase layout in M3D DRAM also creates latency asymmetry across layers, enabling in-memory tiering based on access frequency. In nonvolatile memory (NVM), back end of line compatible, low-temperature processes enable RRAM stacks above CMOS logic; stacking such tiers creates short, wide data paths that enable energy-efficient near-memory computation and high-capacity storage~\cite{du2024monolithic,felfel2020quantifying,pan2025stratum}. CHIME combines the advantages of these two memory devices: M3D DRAM provides fast, durable access for attention and streaming, while M3D RRAM offers dense, low-leakage storage for massive weights.

\subsection{Motivation}
While SOTA PIM acceleration for LLM focus on solving the bandwidth bottleneck with DRAM, on-device MLLMs are still constrained by memory-bound and tight power budgets. These needs must be addressed together: bandwidth for attention and connector kernels, capacity with low leakage for data-intensive KV cache and weights, and endurance aware management for device protection. The system must also preserve digital precision, support variable sequences, and deliver high energy efficiency without rebuilds. M3D DRAM and M3D RRAM provide complementary strengths for this goal. We therefore design a chiplet-based near memory platform, supported by a mapping framework that maps kernels by access locality and bandwidth, tiers the KV cache across M3D DRAM, and migrates data only when reuse outweighs transfer cost. This co-design forms the CHIME system.

\begin{figure}[t]
    \centering
    \includegraphics[width=0.85\linewidth]{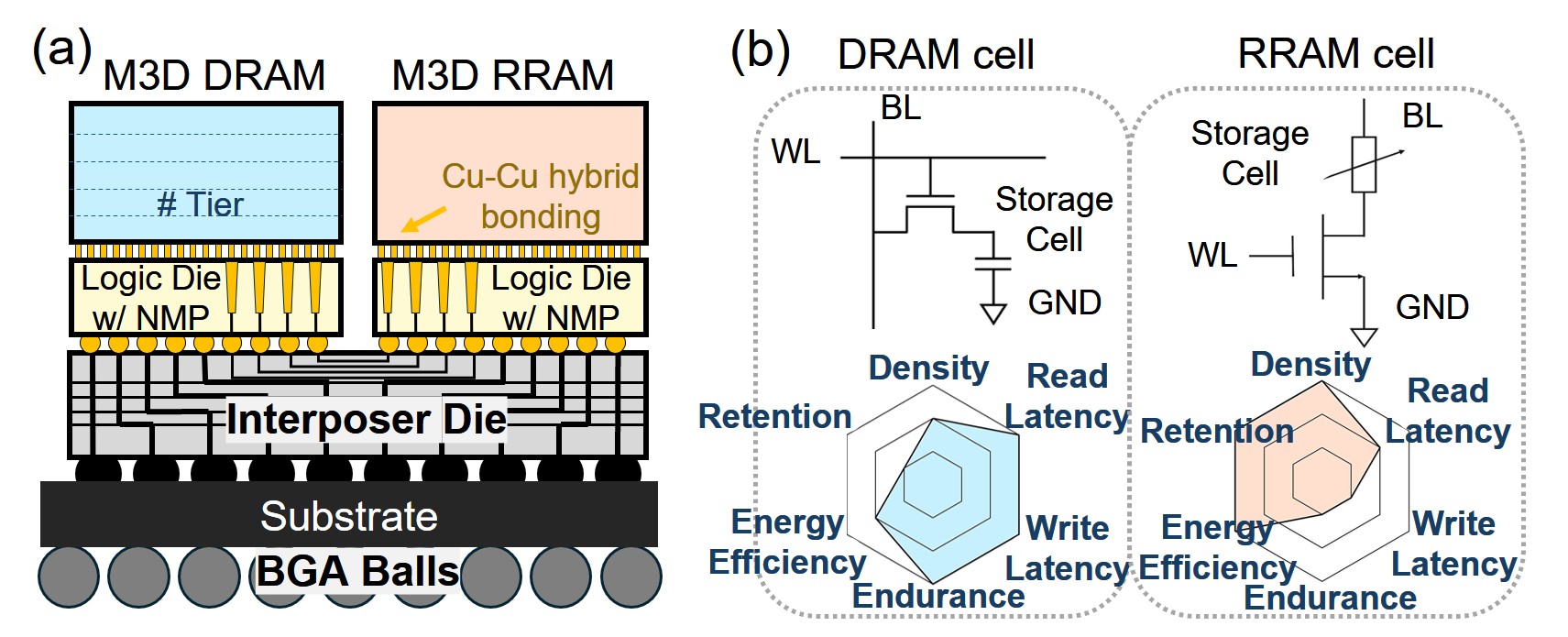}
    \caption{CHIME system overview (a) 2.5D UCIe package integrating M3D DRAM and M3D RRAM on logic dies via interposer (b) Device-level tradeoffs of DRAM (1T1C) and RRAM (1T1R)}
    \label{fig:fig3}
\end{figure}

\section{CHIME Hardware–Software Co-design}

\subsection{CHIME System Overview}
\label{overview}

% \weihong{This is unclear. It is better to first introduce what is the major components of CHIME. Then explain the relationship between the software and hardware components.}

CHIME is a hardware-software co-designed accelerator tailored for edge MLLM inference. Its architecture is built upon two core components: a heterogeneous hardware platform and a co-designed software mapping framework that orchestrates its operation. The hardware platform integrates two distinct near-memory processing chiplets: M3D DRAM and M3D RRAM chiplets in a 2.5D UCIe~\cite{lin202536} package (Fig.~\ref{fig:fig3}(a)), each on its own logic die with near-memory processors (NMPs). Data moves across chiplets via DMA over UCIe, while near-memory kernels operate in place. Our mapping framework exploits this heterogeneity with a hierarchical strategy that places data across chiplets and optimizes local execution.

\begin{itemize}[leftmargin=*]
      \item \textbf{Hardware Design}: Fig.~\ref{fig:fig3}(b) shows the device tradeoffs: DRAM (1T1C) is volatile, fast, and high endurance, and its inherent latency heterogeneity across vertical layers enables in-memory tiering, making it well-suited for latency-critical kernels; RRAM (1T1R) is nonvolatile and dense, well-suited for large model weights, though with higher write energy and limited endurance. The detailed hardware design and implementation are presented in Section~\ref{sec:hardware}.
     \item \textbf{Mapping Framework}: Our framework orchestrates MLLM inference through three core strategies: \encircle{1} workload-aware data layout that statically maps model components to the optimal memory based on MLLM profiling; \encircle{2} KV cache tiered scheduling policy that manages attention KV cache to ensure low-latency access; and \encircle{3} kernel locality-aware fusion that combines operations to maximize data reuse within the near-memory processors, thus minimizing intermediate data movement (Section~\ref{mapping}). 
\end{itemize}

\subsection{Hardware Design and Implementation}\label{sec:hardware}

\subsubsection{M3D DRAM for Latency-critical Computation}

\begin{figure}[t]
    \centering
    \includegraphics[width=1\linewidth]{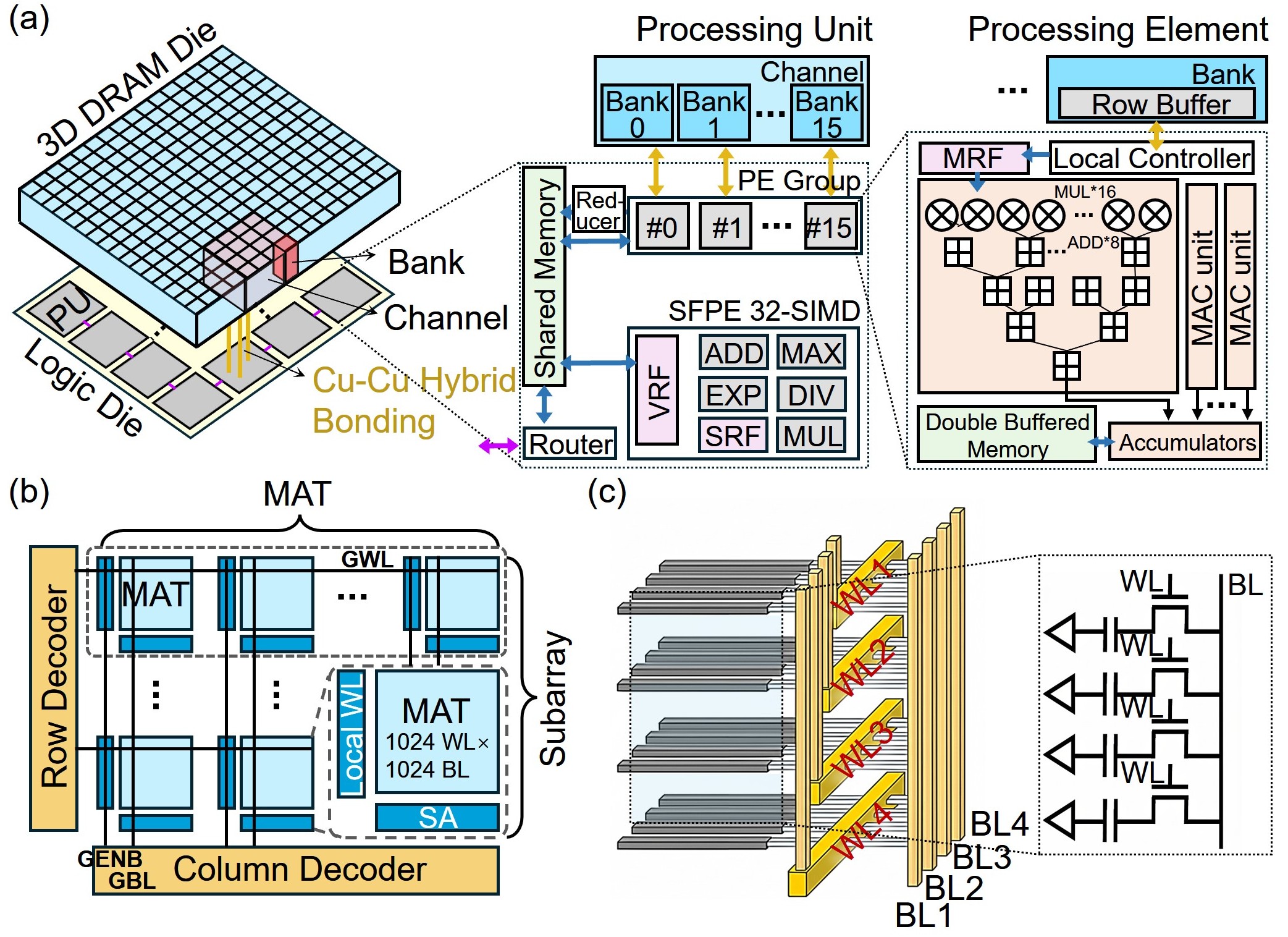}
    \caption{M3D DRAM hardware design (a) M3D DRAM stack with NMP on logic die, organized as channels, banks, and PUs (b) Bank organization with MATs (c) Vertical M3D stacking of 1T1C subarrays}
    \label{fig:fig5}
\end{figure}

The M3D DRAM handles all kernels except the FFN, covering image preprocessing, KV cache, query-key-value (QKV) weight storage, the vision encoder, the connector, and attention. The weight matrices are first placed in DRAM memory, from which data are streamed via MIVs into the logic die for computation. Fig.~\ref{fig:fig5}(a) illustrates the organization: each channel hosts one processing unit (PU) with shared memory for attention activations, a vector register file (VRF), a reducer, a low-latency router, a 256-way SIMD SFPE, and a group of 16 PEs. Each PE includes a matrix register file (MRF), a local controller, double-buffered memory, accumulators, and a $2{\times}2$ MAC tensor core. Double-buffering enables the tensor core to compute on one tile while transferring results from the other, effectively hiding movement latency and maximizing utilization. Activations for attention stay in the local SRAM to avoid costly write-back, while KV cache blocks are written back without frequent overwrites of the same region.

Computation begins as row buffers in each bank stream tiles of QKV to the PU. These tiles pass through the SFPE–PE pipeline, where attention scoring, softmax, and post-softmax scaling are executed directly adjacent to DRAM. Fig.~\ref{fig:fig5}(b) shows how each bank is organized into $1,024\times1,024$ MATs with a 32\,Kb row buffer to supply full rows at channel bandwidth. The vertical integration in Fig.~\ref{fig:fig5}(c) further reduces the distance between data and compute: bitlines (BLs) are stitched across tiers by dense monolithic links, and wordlines (WLs) traverse staircases to expose high internal bandwidth to the PU cluster. To manage the large and growing KV cache, the mapping framework employs tiered placement across the 200-layer M3D stack. Five in-memory tiers place frequently accessed blocks in lower layers, with the hottest attention data in the bottom tier (Tier-0) and connector kernels in the top tier (Tier-4). After attention computation, the attention output (AttnOut) is streamed over UCIe to the M3D RRAM, which executes the FFN kernel and returns the FFN output (FFNOut) to DRAM, enabling the next decoding step without idle cycles. Key M3D DRAM parameters are detailed in Section~\ref{sec:setup}\ref{subsubsec:Hardware Configurations}.

\begin{figure}[t]
    \centering
    \includegraphics[width=1\linewidth]{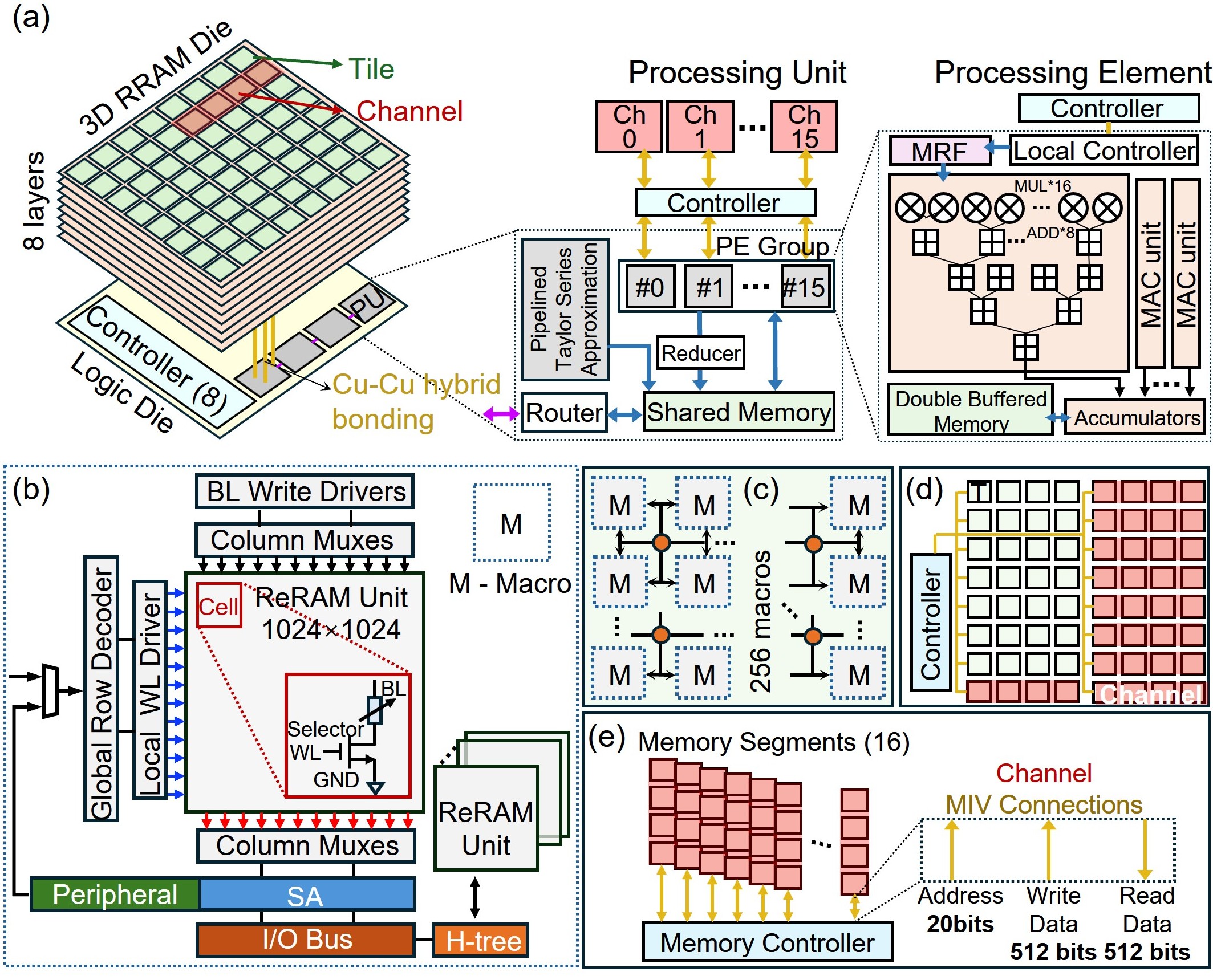}
    \caption{M3D RRAM hardware design (a) 3D RRAM stack with NMP on logic die, organized as channels, controllers, and PUs  (b) RRAM macro (c) Tile organization with local H-trees (d) Die-level layout with controllers, channels, and tiles (e) Memory subsystem with channel and tile I/O}
    \label{fig:fig6}
\end{figure}

\subsubsection{M3D RRAM for Energy-efficient Storage}

The M3D RRAM runs the FFN kernel, where weights are resident in the stacked arrays and later steps access them directly without reload. As shown in Fig.~\ref{fig:fig6}(a), eight RRAM layers sit above the logic die, each managed by a dedicated controller. The FFN input AttnOut arrives from the M3D DRAM and is distributed across 8 controllers to 16 PUs. Each pair of PUs is assigned to one RRAM layer to maximize parallel throughput. Inside each PU, wide data slices are buffered in a 1~MB SRAM that keeps activations local. A lightweight router and reducer feed 16 PEs that perform local MAC operations with weights stored in RRAM. Computation begins from the macro in Fig.~\ref{fig:fig6}(b), a $1{\,}024\times1{\,}024$ RRAM unit with local bitline drivers and column multiplexers. Fig.~\ref{fig:fig6}(c) tiles 256 such macros per tile and links them with 64 local H-trees for synchronous wide reads and writes. The controllers in Fig.~\ref{fig:fig6}(d) schedule accesses across channels and tiles, balancing thermal load and wear while ensuring that each PU receives the required weight and activation data. At the interface in Fig.~\ref{fig:fig6}(e), sixteen memory segments supply a 20 bit address path and 512 bit read and write datapaths that merge at the channel through M3D vertical connections. This pipeline ensures that once AttnOUT enters the logic die, it is immediately fused with preloaded weights, and the resulting FFNOut streams back to the M3D DRAM for the next decoding stage. Key parameters related to M3D RRAM hardware configuration appear in Section~\ref{sec:setup}~\ref{subsubsec:Hardware Configurations}.

\subsection{CHIME Mapping Framework}\label{mapping}
\begin{figure}[t]
    \centering
    \includegraphics[width=1\linewidth]{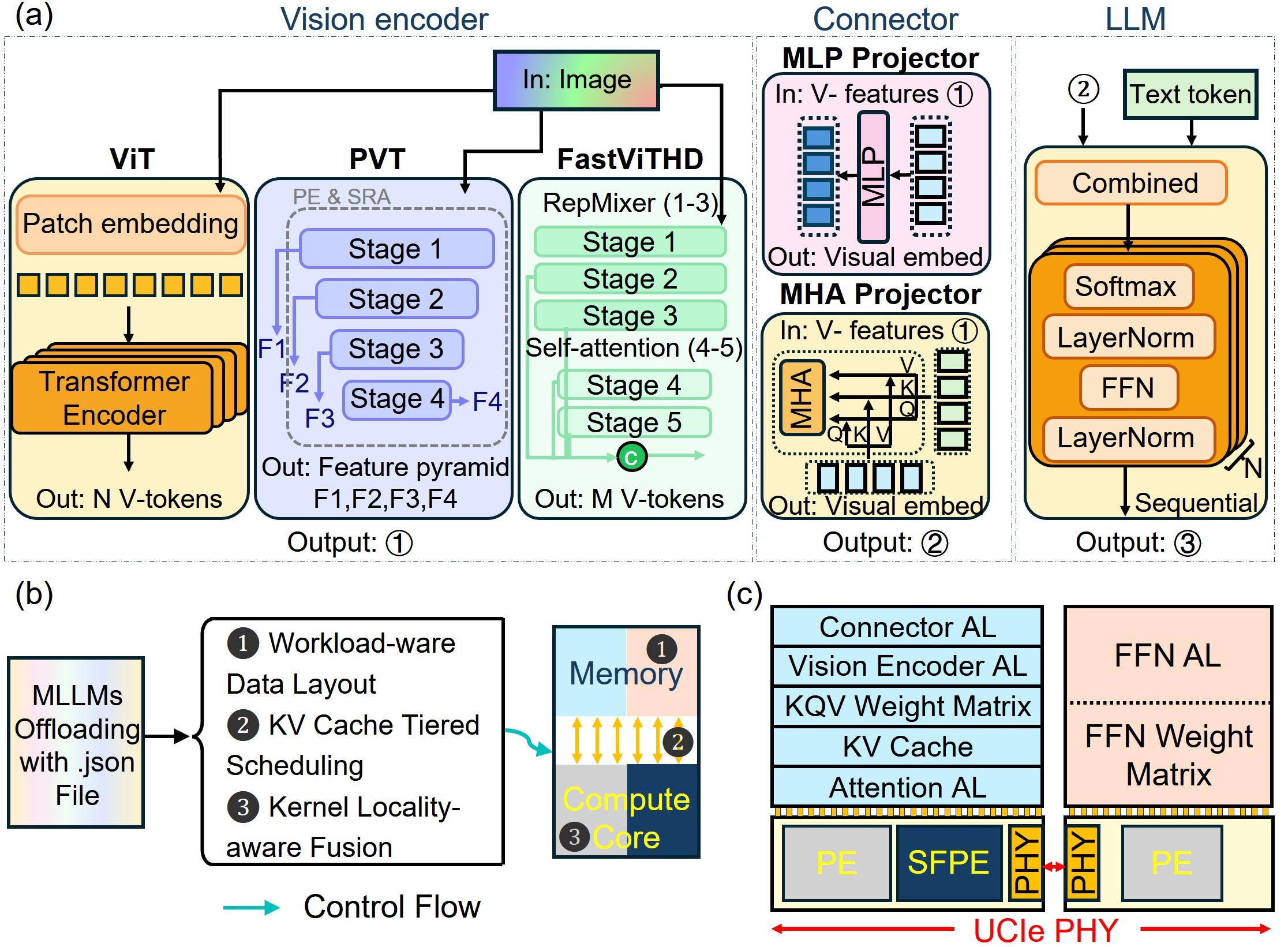}
    \caption{MLLM dataflow and CHIME mapping (a) MLLMs abstraction with vision encoder, connector, and transformer-based LLM (b) Mapping framework pipeline with \encircle{1} workload-aware data layout, \encircle{2} KV cache tiered scheduling, and \encircle{3} kernel locality-aware fusion (c) Mapping implementation}
    \label{fig:fig4}
\end{figure}

The heterogeneous chiplet design of CHIME adds complexity in data management and scheduling. A co-designed mapping framework is required to exploit the complementary strengths of high-bandwidth M3D DRAM and high-density M3D RRAM. This mapping framework deploys general MLLMs onto the CHIME platform by following three design principles that minimize cross-chiplet transfers and keep computation local. Mapping a general MLLM onto this platform first requires a clear understanding of its key components and dataflows. Fig.~\ref{fig:fig4}(a) decomposes a general MLLM into the vision encoder, the connector, and the transformer backbone. The vision encoder transforms an input image into visual tokens and can be implemented as ViT~\cite{dosovitskiy2020image} without downsampling that produces $N$ tokens, PVT~\cite{wang2021pyramid} with a four-stage pyramid downsampling, or FastViT-HD~\cite{vasu2023fastvit} with five-stage downsampling that compresses to $M$ tokens where $M \ll N$. The connector either projects visual features to pseudo tokens via an MLP, or applies cross-attention with visual KV and text Q. The LLM backbone is a transformer with attention, LayerNorm, and FFN, backed by a KV cache that grows with context length. 

As illustrated in Fig.~\ref{fig:fig4}(b), our mapping framework orchestrates the execution of general MLLM components on CHIME heterogeneous platform through three core principles. Fig.~\ref{fig:fig4}(c) summarizes the resulting implementation.

\encircle{1} \textbf{Workload-aware Data Layout.} 
Our data layout is governed by a strict two-cut-point dataflow designed to minimize cross-chiplet traffic between the DRAM-NMP and RRAM-NMP. The DRAM-NMP executes fused kernels for QKV projection and FlashAttention~\cite{dao2022flashattention} streaming attention, while the RRAM-NMP holds large FFN weights and performs a fully fused FFN computation. This partitioning creates two fixed, activation-only transfer points: (1) AttnOut (DRAM$\rightarrow$RRAM) and (2) FFNOut (RRAM$\rightarrow$DRAM). This enables a pipelined execution model: for a given step $t$, the DRAM-NMP computes $AttnOut(t)$ and streams it to the RRAM-NMP for ${FFN}(t)$; the next step ${Attention}(t{+}1)$ can start only after the final ${FFN}(t)$ output is produced. This guarantees that only the small, final AttnOut and FFNOut activations traverse the UCIe interconnect, achieving minimal cross-chiplet traffic.

\encircle{2} \textbf{KV Cache Tiered Scheduling.} 
Our scheduling strategy introduces an endurance-aware KV cache tiering policy that leverages the properties of M3D in-memory tiering. We first exploit the intrinsic vertical latency gradient of M3D-DRAM for intra-stack tiering, placing hotter KV blocks in faster layers (Tier-0) and cooler blocks in slower ones (Tier-1,...Tier-4). For extremely long contexts, the coldest blocks are offloaded to M3D RRAM in a one-shot, write-once manner.

\encircle{3} \textbf{Kernel Locality-aware Fusion.} 
The hardware-aware kernel fusion enables our two-cut-point dataflow, maximizing data locality within each chiplet. Moreover, by utilizing the fast on-die SRAM to temporarily store intermediate data, these fused kernels can complete their operational sequence on the logic die, eliminating costly write-backs to memory. Table~\ref{tab:fusion} lists the fused kernels generated by the mapping framework. On the DRAM-NMP, kernels such as \texttt{FUSED\_QKV\_PROJ} and \texttt{FUSED\_ATTN\_STREAM} avoid materializing the large attention score matrix by passing partial results directly to SFPEs for online softmax and then into subsequent GEMMs within local memory. On the RRAM-NMP, the \texttt{FUSED\_FFN\_ACT} kernel chains two GEMMs to complete the FFN block without offloading the intermediate tensor. The key design principle is that fusion boundaries coincide with chiplet boundaries, never splitting within kernels of the same step.

\begin{table*}[htbp]
\centering
\caption{Fused near-memory kernels}
\label{tab:fusion}
\resizebox{1\textwidth}{!}{
\renewcommand{\arraystretch}{1} % 行距
\begin{tabular}{|l|l|}
\hline
\textbf{Fused PIM kernel} & \textbf{PE and SFPE execution flow} \\ \hline

\textbf{FUSED\_QKV\_PROJ}$(X,W_Q,b_Q,W_K,b_K,W_V,b_V)$ &
PE: GEMM$(X\cdot W_Q)$ $\rightarrow$ SFPE: Add$(b_Q)$ $\rightarrow Q$;
PE: GEMM$(X\cdot W_K)$ $\rightarrow$ SFPE: Add$(b_K)$ $\rightarrow K^\top$;
PE: GEMM$(X\cdot W_V)$ $\rightarrow$ SFPE: Add$(b_V)$ $\rightarrow V$ \\ \hline

\textbf{FUSED\_ATTN\_STREAM}$(Q,K^\top,V,s)$ &
for each tile $(K_t^\top,V_t)$: PE: GEMM$(Q\cdot K_t^\top)$ $\rightarrow$ SFPE: OnlineSoftmaxUpdate $\rightarrow$
PE: GEMM$(\mathrm{Scores}_t\cdot V_t)$ with accumulate $\rightarrow Out$ \\ \hline

\textbf{FUSED\_FFN\_ACT}$(X,W_1,b_1,W_2,b_2)$ &
PE: GEMM$(X\cdot W_1)$ $\rightarrow$ Add$(b_1)$ $\rightarrow$ ACT $\rightarrow$
PE: GEMM$(Y\cdot W_2)$ $\rightarrow$ SFPE: Add$(b_2)$ $\rightarrow Out$ \\ \hline

\textbf{FUSED\_NORM}$(X,g,b)$ &
SFPE: Reduce $\rightarrow$ Normalize $\rightarrow$ Scale $(\times g)$ $\rightarrow$ Shift $(+b)$ $\rightarrow Out$ \\ \hline
\end{tabular}}
\end{table*}

\section{Evaluation}\label{sec:evaluation}

\subsection{Experimental Setup} \label{sec:setup}
\subsubsection{Baselines} We evaluate our CHIME system using two SOTA efficient MLLMs: FastVLM~\cite{vasu2025fastvlm} and MobileVLM~\cite{chu2024mobilevlm}. Table~\ref{tab:mllm-models} lists their model configurations. Inference is benchmarked on a VQA task with a standard input of a $512{\times}512$ astronaut image and $128$ text tokens, producing $488$ output tokens by default unless otherwise specified. We then compare CHIME with a SOTA PIM LLM accelerator FACIL~\cite{seo2025facil} and a representative edge GPU, NVIDIA Jetson Orin NX~\cite{NVIDIA_Jetson_Orin_NX_Datasheet}.

% \begin{table}[t]
% \caption{NVIDIA Jetson Orin NX Specifications}
% \label{tab:orin-nx}
% \renewcommand{\arraystretch}{0.9}
% \centering
% \scalebox{0.9}{
% \begin{tabular}{|l|l|}
% \hline
% \multicolumn{1}{|c|}{\textbf{Feature}} & 
% \multicolumn{1}{c|}{\textbf{Specification}} \\
% \hline
% AI Performance & 157 TOPS \\
% \hline
% GPU & 1024-core NVIDIA Ampere GPU + 32 Tensor Cores \\
% \hline
% GPU Max Freq. & 1.17 GHz \\
% \hline
% CPU & 8-core Arm Cortex-A78AE v8.2 64-bit, 2 MB L2 + 4 MB L3 \\
% \hline
% CPU Max Freq. & 2.0 GHz \\
% \hline
% Memory & 16 GB 256-bit LPDDR5 @ 102.4 GB/s \\
% \hline
% Power Envelope & 10 W – 40 W \\
% \hline
% Core Area & 69.6 mm × 45 mm \\
% \hline
% \end{tabular}
% }
% \end{table}

% \begin{table}[t]
% \caption{NVIDIA Jetson Orin NX Specifications}
% \label{tab:orin-nx}
% \renewcommand{\arraystretch}{0.9}
% \centering
% \scalebox{0.8}{
% \begin{tabular}{|l|l|}
% \hline
% \multicolumn{1}{|c|}{\textbf{Feature}} & 
% \multicolumn{1}{c|}{\textbf{Specification}} \\
% \hline
% AI Performance & 157 TOPS \\
% GPU & 1024-core NVIDIA Ampere GPU + 32 Tensor Cores \\
% GPU Max Freq. & 1.17 GHz \\
% CPU & 8-core Arm Cortex-A78AE v8.2 64-bit, 2 MB L2 + 4 MB L3 \\
% CPU Max Freq. & 2.0 GHz \\
% Memory & 16 GB 256-bit LPDDR5 @ 102.4 GB/s \\
% Power Envelope & 10 W – 40 W \\
% Core Area & 69.6 mm × 45 mm \\
% \hline
% \end{tabular}
% }
% \end{table}

\begin{table}[ht] 
\caption{MLLM Model Configurations for Evaluation} 
\label{tab:mllm-models} 
\renewcommand{\arraystretch}{1.05} 
\centering 
\scalebox{0.8}{ 
    \begin{tabular}{|l|l|l|c|} 
        \hline \textbf{Model} & \textbf{Vision Encoder} & \textbf{Connector} & \textbf{LLM Parameters} \\ 
        \hline FastVLM (0.6B) & FastViTHD & Lightweight MLP & Qwen2-0.5B \\ 
        FastVLM (1.7B) & FastViTHD & Lightweight MLP & Qwen2-1.5B \\ 
        MobileVLM (1.7B) & ViT & LDP & LLaMA-1.4B \\ 
        MobileVLM (3B) & ViT & LDP & LLaMA-2.7B \\ 
        \hline 
    \end{tabular} } 
\end{table}

\subsubsection{Hardware Configurations}\label{subsubsec:Hardware Configurations}

CHIME co-packages two near-memory stacks with complementary roles, a 200-layer M3D DRAM (five-tier) optimized for latency-critical computation and an 8-layer M3D RRAM optimized for energy-efficient storage. Table~\ref{tab:dram} reports the device, system, and NMP parameters of the M3D DRAM stack, and Table~\ref{tab:rram} reports the corresponding parameters of the M3D RRAM stack.

\subsubsection{Evaluation Platform}We develop an in house simulator for the CHIME system. We model the 3D RRAM arrays and peripheral circuits with NeuroSim~\cite{peng2020dnn+} to ensure device-level accuracy, and we synthesize the Register Transfer Level (RTL) design in SystemVerilog using Synopsys Design Compiler with a 40\,nm CMOS PDK at 1\,GHz. The synthesis included the tensor core PE, SFPE, and the on-chip controller. All results are scaled to 7nm technology node using established models~\cite{stillmaker2017scaling}.

\begin{table}[t]
    \centering
    \caption{Hardware configurations of 3D RRAM}
    \label{tab:rram}
    \scalebox{0.8}{
        \begin{tabular}{|l|c|l|c|}
            \hline
            \multicolumn{4}{|c|}{\textbf{3D RRAM Device Parameters}} \\ 
            \hline
            \#Layer & 8 & Technology Node & 28 nm CNFET \\ 
            \hline
            Unit Size & 1k$\times$1k & Unit/Tile & 256 \\ 
            % \hline
            % Die Area & 33.6 mm\textsuperscript{2} & Density & 487.6 bit/\textmu m\textsuperscript{2} \\
            \hline
            Read Latency & 2.3 ns & Write Latency & 11 ns \\
            \hline
            Read Energy & 0.4 pJ/bit & Write Energy & 1.33 pJ/bit \\
            \hline
            % Clock Frequency & 1.5 GHz & Avg. Power & 9.5 W/cm\textsuperscript{2} \\
            \multicolumn{4}{|c|}{\textbf{3D RRAM System Parameters}} \\
            \hline
            Chip Capacity & 2 GB & Internal Parallelism & 128 channels \\
            \hline
            Organization & \multicolumn{3}{l|}{\begin{tabular}[l]{@{}l@{}}8 controllers; 16 channels/controller; 4 tiles/channel\end{tabular}} \\
            \hline
            Interconnect & \multicolumn{3}{l|}{64 H-trees connect 256 units within a tile} \\
            \hline
            Interface BW & \multicolumn{3}{l|}{Peak BW = 512 GB/s (8 controllers $\times$ 512 bit $\times$ 1 GHz)} \\
            \hline
            \multicolumn{4}{|c|}{\textbf{Processing Element (PE)}} \\
            \hline
            Tensor Core & 4$\times$4 MACs & Double Buffered SRAM & 8 KB \\ 
            \hline
            \multicolumn{4}{|c|}{\textbf{Processing Unit (PU)}} \\
            \hline
            \#PEs & 16 & Shared Memory & 80 KB \\
            \hline
            Special Function PE & None & Ring Router & 128 GB/s/link \\
            \hline
            \multicolumn{4}{|c|}{\textbf{RRAM NMP Processor}} \\
            \hline
            Basic & \multicolumn{3}{l|}{7 nm process; 0.7 V supply; 33.6 mm\textsuperscript{2} die area; FP16 format.} \\
            \hline
            \#PUs & 16 & SRAM Capacity & 1 MB \\
            \hline
            Peak Performance & 32 TFLOPS & Peak Power & 2.584 W \\ 
            \hline
        \end{tabular}
    }
\end{table}

\begin{table}[t]
    \centering
    \caption{Hardware configurations of 3D DRAM}
    \label{tab:dram}
    \scalebox{0.8}{
        \begin{tabular}{|l|c|l|c|}
            \hline
            \multicolumn{4}{|c|}{\textbf{3D DRAM Device Parameters}} \\ 
            \hline
            \#layers & 200 & Technology Node & 35 nm \\
            %\hline
            %BL/WL Pitch & 70 nm/1 um & Staircase Pitch & 500 nm \\
            \hline
            MAT Size & 1k$\times$1k & \#MATs/Bank & 200 \\
            \hline
            Bank Capacity & 200 Mb & Bank Area & 0.439 mm\textsuperscript{2} \\
            \hline
            Row Buffer & 32 Kb & Read/Write Energy/bit & 0.429 pJ \\
            \hline
            Chip Area & 121 mm\textsuperscript{2} & Read/Write Latency/ns & (3+0.8*L) ns \\
            \hline
            \multicolumn{4}{|c|}{\textbf{3D DRAM System Parameters}} \\
            \hline
            Tier Design & \multicolumn{3}{l|}{5 tiers (L1, L2, L3, L4, L5); 1.25 GB capacity per tier.} \\
            \hline
            Organization & \multicolumn{3}{l|}{\begin{tabular}[l]{@{}l@{}}16 channels per chip (64b data I/O per channel);\\ 16 banks per channel.\end{tabular}} \\
            % \hline
            % DRAM Timing & \multicolumn{3}{l|}{\begin{tabular}[l]{@{}l@{}}tRCD=[1.10, 1.46, 1.89, 2.29, 3.82] ns; \\ tRP=4.77 ns; tRAS=tRCD+3.82ns; tRC=tRP+tRAS.\end{tabular}} \\
            % \hline
            % xPU-DRAM I/F & \multicolumn{3}{l|}{1024b data I/Os; 6.4 Gbps per pin (same as HBM3)} \\
            \hline
            \multicolumn{4}{|c|}{\textbf{Processing Element (PE)}} \\
            \hline
            Tensor Core & 2$\times$2 MACs & Double Buffered SRAM & 1 KB \\% Psum 
            % \hline 
            % Tiering Table & 8$\times$8b Registers & Row Swap Buffer & 8KB RF \\
            % \hline
            % Tiering Table & \multicolumn{3}{l|}{8$\times$8b Registers} \\
            \hline
            \multicolumn{4}{|c|}{\textbf{Processing Unit (PU)}} \\
            \hline
            \#PEs & 16 & Shared Memory & 20 KB \\
            \hline
            Special Function PE & 256-way SIMD & Ring Router & 128 GB/s/link \\
            \hline
            \multicolumn{4}{|c|}{\textbf{DRAM NMP Processor}} \\
            \hline
            Basic & \multicolumn{3}{l|}{7 nm process; 0.7 V supply; 121 mm\textsuperscript{2} die area; FP16 format.} \\
            \hline
            \#PUs & 16 & SRAM Capacity & 512 Kb \\
            \hline
            Peak Performance & 2 TFLOPS & Peak Power & 0.671 W \\
            % \hline
            % \begin{tabular}[c]{@{}l@{}}Aggregated On-chip\\ Ring Bandwidth\end{tabular} & 2.048 TB/s & \begin{tabular}[c]{@{}l@{}}Aggregated Mono3D\\ SRAM Capacity\end{tabular} & 36 Mb \\
            \hline
        \end{tabular}
    }
\end{table}

\subsection{Speedup and Energy Efficiency Analysis}
\begin{figure}[t]
    \centering
    \includegraphics[width=1\linewidth]{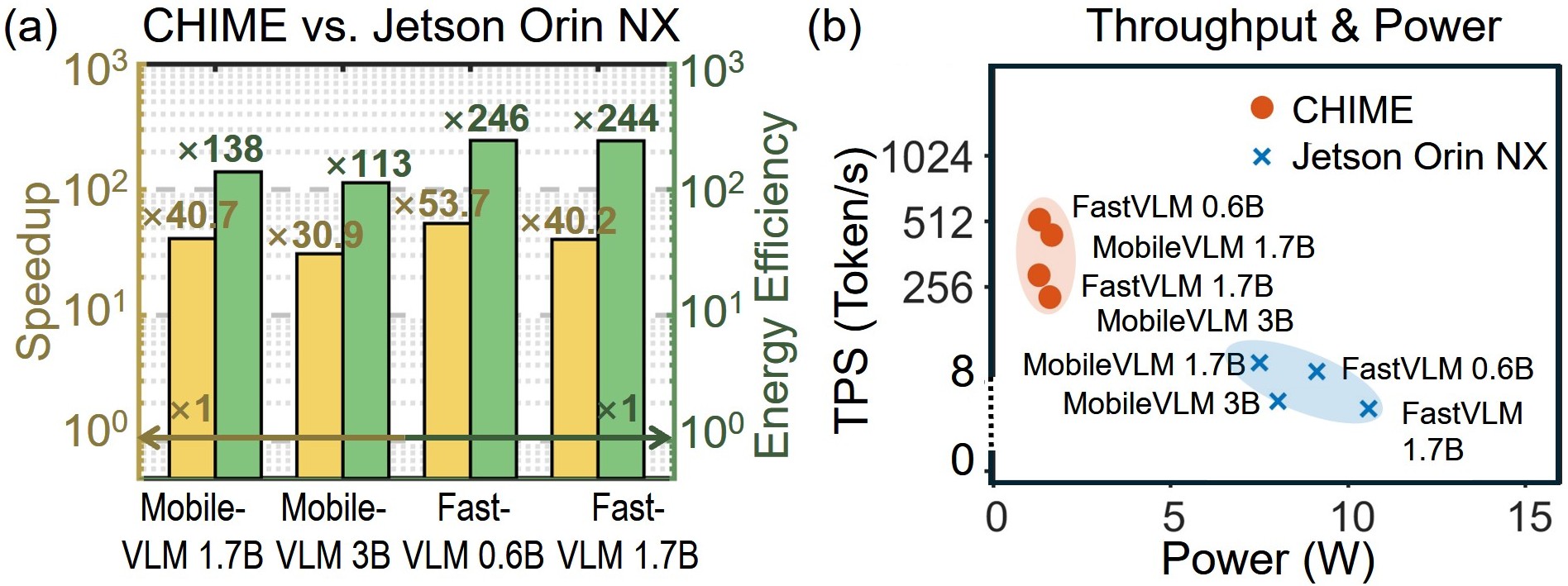}
    \caption{Performance comparison over Jetson Orin NX~\cite{NVIDIA_Jetson_Orin_NX_Datasheet} baseline across MobileVLM~\cite{chu2024mobilevlm} and FastVLM~\cite{vasu2025fastvlm} models (a) Speedup and energy efficiency comparison (b) Throughput and power comparison}
    \label{fig:fig7}
\end{figure}

Fig.~\ref{fig:fig7}(a) compares CHIME with Jetson Orin NX~\cite{NVIDIA_Jetson_Orin_NX_Datasheet} on MobileVLM~\cite{chu2024mobilevlm} and FastVLM~\cite{vasu2025fastvlm}. CHIME achieves a $\sim$41$\times$ speedup (arithmetic-mean; $31$–$54\times$ across models) and an energy-efficiency gain of $\sim$185$\times$ (arithmetic-mean; $113$–$246\times$ across models). The gains are larger for the smaller variants in each family (MobileVLM~1.7B vs.~3B; FastVLM~0.6B vs.~1.7B). Their working dimension fits far more fully in the near-memory tiers with less on chip data movement, sustaining PE - SFPE pipelines and reducing UCIe data travel, while the baseline remains limited by off-chip DRAM traffic. Fig.~\ref{fig:fig7}(b) illustrates throughput, measured in tokens per second (TPS), and power. Jetson draws $7–11$\,TPS at $7-13$\,W while CHIME delivers $233–533$\,TPS at around $2$\,W, yielding roughly $40\times$ higher throughput at $5\times$ lower power. This outcome reveals the core inefficiency of the Jetson GPU for this workload: its power budget is spent on a compute engine largely stalled by memory access, whereas CHIME's co-design invests its tight budget power in high-bandwidth data movement, proving a fundamentally more effective architectural tradeoff. 

We then compare CHIME with a SOTA PIM LLM accelerator FACIL~\cite{seo2025facil} and the representative edge GPU, NVIDIA Jetson Orin NX~\cite{NVIDIA_Jetson_Orin_NX_Datasheet} (Table~\ref{tab:acc-specs}). The comparison with the SOTA PIM accelerator FACIL further underscores CHIME’s advantages (see Table~\ref{tab:acc-specs}). While FACIL reaches $19.3$ token/s at $38.5$ W, CHIME’s throughput of $233-533$ token/s represents a $12.1\times$ to $69.2\times$ throughput leap at a lower power envelope of $\sim2$ W. This significant improvement highlights the high performance of CHIME’s heterogeneous memory architecture and co-designed mapping framework in alleviating the memory bottleneck more effectively than other PIM-based approaches.

\begin{table}[t]
 \caption{Performance Comparison of Edge AI Platforms \\ }

\label{tab:acc-specs}
\centering
\scalebox{0.78}{
\begin{tabular}{|l|c|c|c|}
\hline
\textbf{Specification} 
& \textbf{Jetson Orin NX~\cite{NVIDIA_Jetson_Orin_NX_Datasheet}} 
& \textbf{FACIL~\cite{seo2025facil}} 
& \textbf{CHIME} \\
\hline
Design & GPU & Near-bank DRAM & Our work \\
Technology Node (nm) & 8 & 15 &  28\&35 \\
Frequency (GHz) & $\le$0.92 & $\le$3.2 & 1 \\
Die Area (mm$^2$) & $\sim$200 & $\sim$200 & 28.71\&24.85 \\
Power (W) & 10-40 & 5.7-38.5  & 2 \\
Throughput (Token/s) & 7.4-11 & 7.7-19.3 & 233-533 \\
\hline
Energy Eff. (Token/J) & 0.28-0.74 & 0.50-1.35 & 116.5-266.5 \\
Hardware Eff. (Token/s/mm$^2$) & 0.037-0.055 & 0.039-0.097& 4.35-9.95 \\
\hline
\end{tabular}
}
\end{table}

% \begin{table}[t]
% \caption{Comparison with State-of-the-art MLLM Accelerators}
% \label{tab:acc-specs}
% \renewcommand{\arraystretch}{0.95}
% \centering
% \scalebox{0.78}{
% \begin{tabular}{|l|c|c|c|}
% \hline
% \textbf{Specification} 
% & \textbf{StreamDCIM~\cite{qin2025streamdcim}} 
% & \textbf{MulTCIM~\cite{tu2023multcim}} 
% & \textbf{CHIME} \\
% \hline
% Technology Node (nm) & $28$ & $28$ &  $25\&35$ \\
% Frequency (GHz) & $0.2$ & $0.085-0.275$ & $1$ \\
% Die Area (mm$^2$) & $12.10$ & $14.36$ & $28.71\&24.85$ \\
% Power (W) & $0.12$ & $0.029-0.15$ & $1.93-2.92$ \\
% Throughput (TOPS) & $2.40$ & $2.49-12.86$ &  \\
% \hline
% Energy Eff. (TOPS/W) & &  & \\
% Hardware Eff. (TOPS/mm$^2$) & & &  \\
% \hline
% \end{tabular}
% }
% \end{table}

% \begin{table}[t]
%   \centering
%   \caption{Tokens per second (TPS) comparison}
%   \label{tab:tps}
%   \scalebox{0.80}{
%     \begin{tabular}{|l|c|c|c|c|}
%       \hline
%      \textbf{Token/s} & \textbf{MobileVLM 1.7B} & \textbf{MobileVLM 3B} & \textbf{FastVLM 0.5B} & \textbf{FastVLM 1.5B} \\
%       \hline
%       \textbf{Jetson} & 10.95 & 7.54 & 9.93 & 7.40 \\
%       \hline
%       \textbf{CHIME}       & 411.73 & 212.7207 & 484.70 & 268.604 \\
%       \hline
%     \end{tabular}
%   }
% \end{table}

\subsection{Area and Power Overhead}
\begin{figure}[t]
    \centering
    \includegraphics[width=1\linewidth]{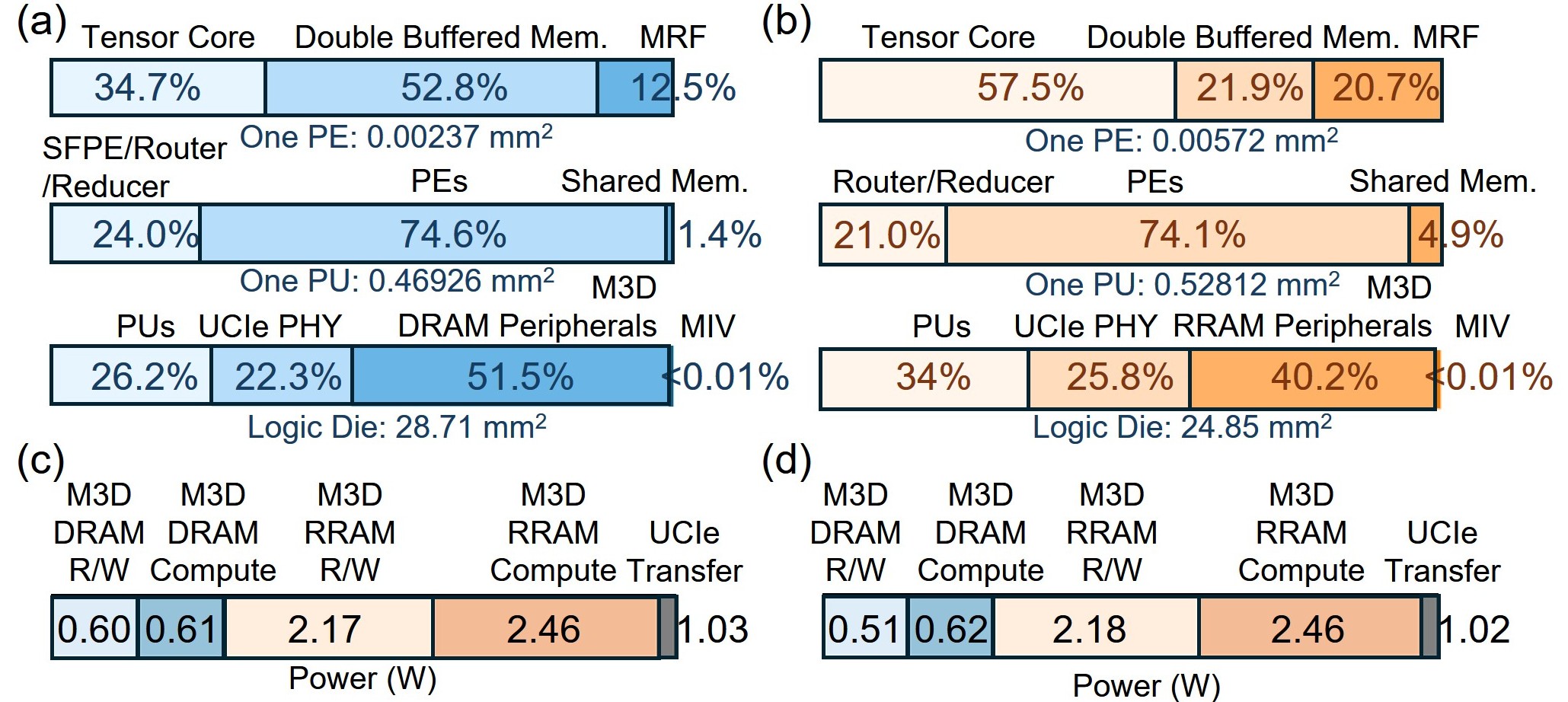}
    \caption{CHIME logic-die area and power breakdown (a) Area breakdown of M3D DRAM (b) Area breakdown of M3D RRAM (c) Power breakdown of FastVLM (0.6B) (d) Power breakdown of MobileVLM (1.7B)}
    \label{fig:fig8}
\end{figure}

Fig.~\ref{fig:fig8}(a) and (b) show the logic die area breakdown. In M3D DRAM, peripherals use $51.5$\%, the UCIe PHY $22.3$\%, and PUs $26.2$\%. In M3D RRAM, larger tensor cores and double buffered SRAM raise the PU share to $34.0$\%. The total logic die area is smaller, $24.85$\,mm$^2$ versus $28.71$\,mm$^2$, with lower peripheral cost. The split matches roles: DRAM PEs emphasize buffering for attention and connector kernels. RRAM PEs devote more compute and registers for high throughput FFN streaming. Fig.~\ref{fig:fig8}(c) and (d) show the power breakdown. RRAM dominates because it runs the data-intensive FFN. DRAM runs attention at lower power. Power stays stable across models, which implies utilization drives power more than model size. The UCIe link draws about $1$\,W.

\begin{figure}[t]
    \centering
    \includegraphics[width=1\linewidth]{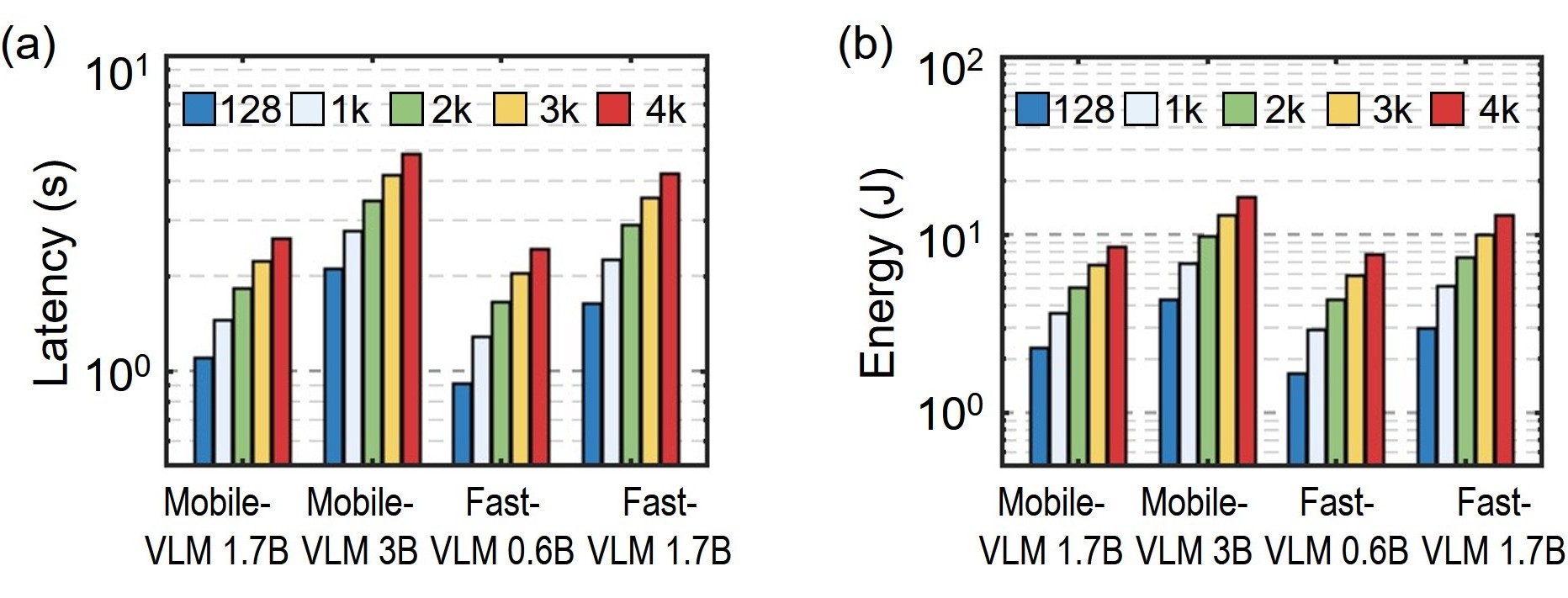}
    \caption{Impact of sequence length: from 128 to 4k (a) Latency vs. sequence length (b) Energy vs. sequence length}
    \label{fig:fig9}
\end{figure}

\begin{figure}[!ht]
    \centering
    \includegraphics[width=1\linewidth]{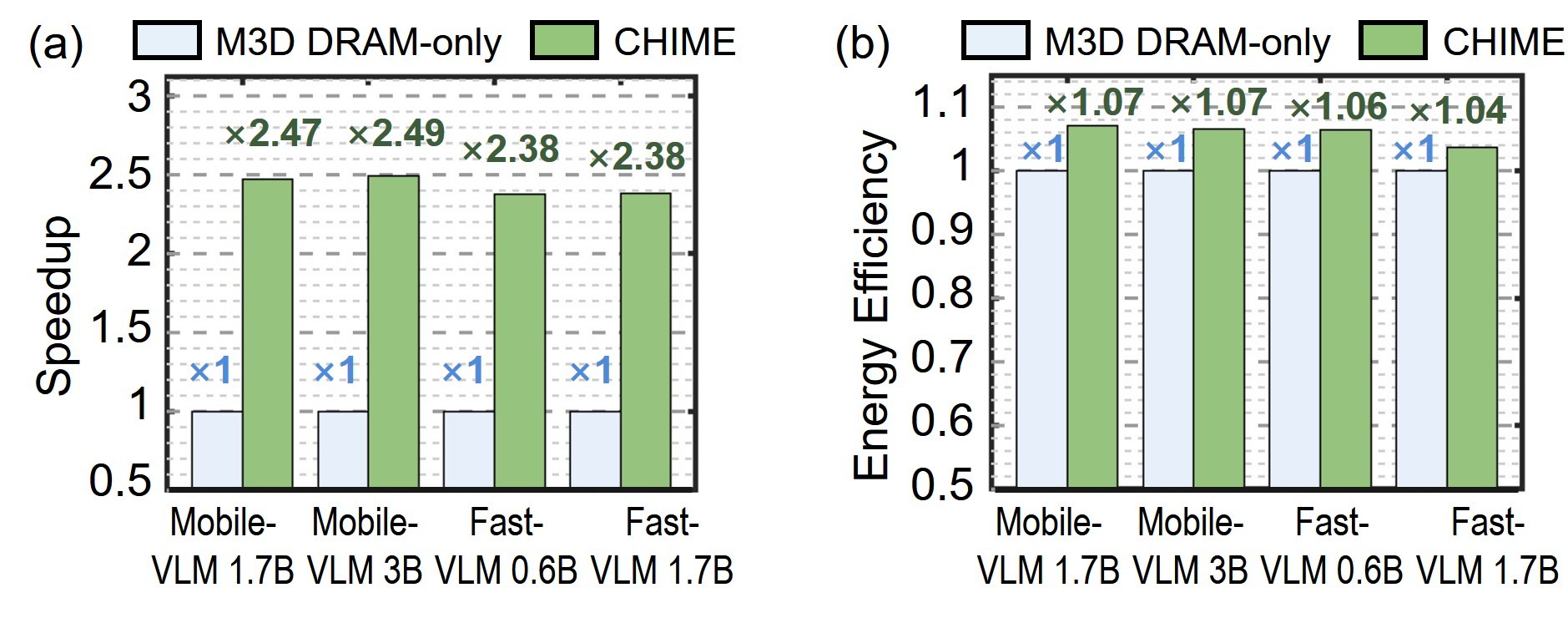}
    \caption{Impact of memory configuration: CHIME vs. M3D DRAM-only (a) Speedup comparison (b) Energy efficiency comparison}
    \label{fig:fig10}
\end{figure}

\subsection{Sensitivity Analysis}
Real deployments face wide variability in prompt length and memory hierarchy choices, both directly determine KV-cache traffic, bandwidth pressure, and thus performance on edge devices. We therefore study two sensitivities of CHIME: input sequence length and memory configuration, to expose the interaction between workload scaling and memory organization.

\subsubsection{Sequence Length} Fig.~\ref{fig:fig9}(a) and (b) show accelerator-centric latency (end-to-end delay in ms) and total energy per inference (in Joules) as the text length grows from 128 to 4k tokens for MobileVLM~1.7B/3B and FastVLM~0.6B/1.7B. Across models, both metrics increase almost linearly with length (roughly an order-of-magnitude from 128 to 4k), consistent with decoding that streams the KV cache every step. Larger models exhibit steeper slopes: MobileVLM~3B and FastVLM~1.7B incur the higher latency and energy due to wider QKV/FFN matrices and a larger KV footprint per token, while FastVLM~0.6B is lowest. At short contexts the gaps narrow as vision encoding and connector dominate, while at long contexts decoding dominates and model size is the main driver.

\subsubsection{Memory Configuration} Fig.~\ref{fig:fig10} validates the heterogeneous architecture by comparing CHIME to a DRAM-only baseline, demonstrating speedups of $2.38–2.49\times$ and energy efficiency gains of $1.04–1.07\times$. The speedup is significant for larger models, especially MobileVLM~3B whose FFN weights overwhelm DRAM-centric M3D DRAM. By offloading these capacity-oriented weights to RRAM, CHIME preserves critical DRAM bandwidth and fast access capability for latency-critical attention kernels, keeping the PE - SFPE pipeline fed to sustain higher throughput.

\section{Conclusion}
Edge MLLM inference is constrained by memory bottlenecks, where bandwidth and data intensity drive high power consumption. Our proposed CHIME addresses this challenge with a 2.5D near-memory architecture pairing M3D DRAM and RRAM. DRAM serves latency-critical kernels through tiered KV placement, while RRAM stores dense FFN weights, exploiting its non-volatility and low leakage. A workload-aware mapping framework optimizes data layout, KV scheduling, and kernel fusion to improve locality, raising effective bandwidth and reducing data movement.
On FastVLM and MobileVLM, CHIME achieves up to $41.4\times$ speedup and $185.3\times$ energy efficiency over Jetson Orin NX GPU, sustaining $116.5–266.5$ token/J versus $0.7–1.1$ token/J, and outperforms FACIL by $12.1–69.2\times$ in TPS. Compared with an M3D DRAM–only design, heterogeneous memory delivers $2.4\times$ performance and $7$\% energy gains, validating that reserving DRAM bandwidth for attention while offloading weights to RRAM is an effective edge design.

\section*{Acknowledgment}
ChatGPT-5 was used to check grammar and enhance clarity of the text.
This work was supported in part by PRISM and CoCoSys—centersand CoCoSys—centers in JUMP 2.0, an SRC program sponsored by DARPA (SRC grant number - 2023-JU-3135). This work was also supported by NSF grants \#2003279, \#1911095, \#2112167, \#2052809, \#2112665, \#2120019, \#2211386.

\balance 

\bibliographystyle{IEEEtran}
\bibliography{Ref}

\end{document}